# NEW DATA SECURITY REQUIREMENTS AND THE PROCEDURALIZATION OF MASS SURVEILLANCE LAW AFTER THE EUROPEAN DATA RETENTION CASE

Frederik Zuiderveen Borgesius[1] & Axel Arnbak[2]

[1]Dr. Frederik Zuiderveen Borgesius, researcher at the Institute for Information Law (IViR), University of Amsterdam, The Netherlands, f.j.zuiderveenborgesius@uva.nl.
[2]Axel Arnbak, researcher at the Institute for Information Law (IViR), University of Amsterdam, The Netherlands, Research Affiliate, Berkman Center for Internet & Society, Harvard University, A.M.Arnbak@uva.nl.
This paper builds on, and includes sentences from, the PhD research of both authors. Mainly Section II builds on earlier work by Zuiderveen Borgesius. (See F.J. Zuiderveen Borgesius, Improving privacy Protection in the Area of Behavioural Targeting, Kluwer Law International 2015, chapter 3 and 4). Mainly Section V builds on work by Arnbak. (See A.M. Arnbak, Securing Free Private Communications, Forthcoming 2015, chapter 4).

**Abstract** - This paper discusses the regulation of mass metadata surveillance in Europe through the lens of the landmark judgment in which the Court of Justice of the European Union struck down the Data Retention Directive. The controversial directive obliged telecom and Internet access providers in Europe to retain metadata of all their customers for intelligence and law enforcement purposes, for a period of up to two years. In the ruling, the Court declared the directive in violation of the human rights to privacy and data protection. The Court also confirmed that the mere collection of metadata interferes with the human right to privacy. In addition, the Court developed three new criteria for assessing the level of data security required from a human rights perspective: security measures should take into account the risk of unlawful access to data, and the data's quantity and sensitivity. While organizations that campaigned against the directive have welcomed the ruling, we warn for the risk of proceduralization of mass surveillance law. The Court did not fully condemn mass surveillance that relies on metadata, but left open the possibility of mass surveillance if policymakers lay down sufficient procedural safeguards. Such proceduralization brings systematic risks for human rights. Government agencies, with ample resources, can design complicated systems of procedural oversight for mass surveillance – and claim that mass surveillance is lawful, even if it affects millions of innocent people.

# Table of Contents

# I    INTRODUCTION

In 2014 the Court of Justice of the European Union invalidated the Data Retention Directive.[3] The directive obliged all telecom providers in the European Union to retain metadata of all their customers for intelligence and law enforcement purposes, for a period of up to two years. This paper focuses on the following question. What are the implications of the Data Retention judgment for data collection as an interference with privacy, for data security requirements, and for the regulation of mass surveillance law in Europe?

We begin with an introduction to the right to privacy and the right to the protection of personal data in Europe (Part II). Next, we summarize the main elements of the Data Retention Directive, and the Data Retention judgment (Part III). While we do not aim to conduct comparative research in this paper, we do highlight some salient points where EU and US law differ.[4] Our aim with this section is to explain to an international readership the complicated European constitutional situation, with its two institutional systems (the Council of Europe and the European Union) and its two constitutional courts.

Then the paper takes a more analytical turn. We focus on three themes: (i) data collection as a privacy interference, (ii) data security requirements, and (iii) the proceduralization of mass surveillance law. Part IV discusses criteria for metadata collection and access to such data after the judgment. The Court of Justice of the European Union reaffirms that collecting metadata about people interferes with privacy, regardless of how the data are used.

Part V discusses the judgment's implications for data security. The Court develops three new criteria for assessing the level of data security needed from a human rights perspective. Data security measures should take into account the quantity and the sensitivity of data, and the risk of unlawful access to data. The ruling will have a large impact beyond the regulation of surveillance, as it develops a completely new framework for the regulation of data security in the EU The Court substantially raises the obligations for the EU lawmaker to protect data security in legislation.

---

[3] CJEU, C-293/12 and C-594/12, *Digital Rights Ireland and Seitlinger a.o.*, 8 April 2014.
[4] See generally on the difference between EU and US regulation of mass surveillance: Joel R. Reidenberg, *The Data Surveillance State in the United States and Europe,* 49 WAKE FOREST L.REV. 583 (2014).

In Part VI we warn against proceduralization of surveillance law. The Court did not fully condemn mass surveillance that relies on metadata. Rather, the Court set out procedural conditions that must be met to conduct mass surveillance in compliance with the law. It seems that, as long as the procedures are taken care of, far-reaching privacy violations can be claimed to be lawful. In Part VII we discuss the road ahead for mass data surveillance in Europe, and Part VIII concludes.

## II    PRIVACY AND DATA PROTECTION RIGHTS IN EUROPE

In this Part we give an introduction to the right to privacy and the right to the protection of personal data in Europe. First we introduce the European Convention on Human Rights and the European Court of Human Rights, both connected to the Council of Europe (Section II A and II B). Then we turn to the European Union: we discuss the Charter of Fundamental Rights of the European Union, the EU data privacy directives, and the Court of Justice of the European Union (Section II C – II E).

### A.  *European Convention on Human Rights*

The right to privacy is an internationally recognized human right, and is included in the International Covenant on Civil and Political Rights.[5] In Europe, the right to privacy is laid down in the European Convention on Human Rights, a treaty of the Council of Europe that entered into force in 1953.[6] The Council of Europe is the most important human rights organization in Europe. It is based in Strasburg and has 47

---

[5] International Covenant on Civil and Political Rights art. 17, Dec. 16, 1966, S. Treaty Doc. No. 95-20, 6 I.L.M. 368 (1967), 999 U.N.T.S. 171. The Universal Declaration of Human Rights from 1948 contains a provision that protects privacy in Article 12 (Universal Declaration of Human Rights art. 12, G.A. Res. 217A (III), U.N. Doc. A/810 at 71 (1948)).
[6] Convention for the Protection of Human Rights and Fundamental Freedoms, art. 8, Nov. 4, 1950, 213 U.N.T.S. 222.

Member States, including all European Union Member States.[7] All Council of Europe Member States have signed up to the European Convention on Human Rights.[8] Article 8 of the European Convention on Human Rights protects the right to respect for private life. In this paper, we use "privacy" and "private life" interchangeably.[9] Article 8 reads as follows.

> European Convention on Human Rights
>
> Article 8, right to respect for private and family life
>
> 1. Everyone has the right to respect for his private and family life, his home and his correspondence
>
> 2. There shall be no interference by a public authority with the exercise of this right except such as is in accordance with the law and is necessary in a democratic society in the interests of national security, public safety or the economic well-being of the country, for the prevention of disorder or crime, for the protection of health or morals, or for the protection of the rights and freedoms of others.

In brief, article 8 prohibits privacy interferences. Paragraph 2 shows that this prohibition isn't absolute; in many circumstances the state can limit the right to privacy in the view of other interests, such as crime prevention or national security. Article 8 has been applied in several cases on surveillance, as discussed in the next section.

---

[7] Council of Europe, Who We Are, www.coe.int/en/web/about-us/who-we-are, accessed 25 July 2015.
[8] Council of Europe, Country Profiles, www.coe.int/en/web/portal/country-profiles, accessed 25 July 2015.
[9] On the difference between "privacy" and "private life", see: GLORIA GONZÁLEZ FUSTER, *THE EMERGENCE OF PERSONAL DATA PROTECTION AS A FUNDAMENTAL RIGHT OF THE EU (PhD thesis Free University of Brussels)* (Springer 2014), p. 82-84; p. 255.

### B. European Court of Human Rights

The European Court of Human Rights, based in Strasburg, rules on alleged violations of the rights in the European Convention on Human Rights.[10] The European Court of Human Rights interprets the Convention's right to privacy generously, and refuses to pin itself down to one definition. The Court "does not consider it possible or necessary to attempt an exhaustive definition of the notion of private life."[11] The Court says it takes "a pragmatic, common-sense approach rather than a formalistic or purely legal one."[12] This approach allows the Court to adapt the privacy protection of article 8 to new circumstances. The Court says "the Convention is a living instrument which must be interpreted in the light of present-day conditions."[13] The Court uses a "dynamic and evolutive" interpretation of the Convention,[14] and states that "the term 'private life' must not be interpreted restrictively:"[15]

> It is of crucial importance that the Convention is interpreted and applied in a manner which renders its rights practical and effective, not theoretical and illusory. A failure by the Court to maintain a dynamic and evolutive approach would indeed risk rendering it a bar to reform or improvement (…).[16]

This living instrument doctrine allows the Court to apply the right to privacy in unforeseen situations and to new developments.[17] The living instrument doctrine could be seen as the opposite of originalism, one of the major constitutional interpretation principles in US constitutional law.[18]

---

[10] Article 19 and 34 of the European Convention on Human Rights.

[11] See e.g. ECtHR, Niemietz v. Germany, No. 13710/88, 16 December 1992, par. 29. The Court consistently confirms this approach. See e.g. ECtHR, Pretty v. United Kingdom, No. 2346/02, 29 April 2002, par. 61; ECtHR, S. and Marper v. United Kingdom, No. 30562/04 and 30566/04. 4 December 2008, par. 66.

[12] ECtHR, Botta v. Italy (153/1996/772/973), 24 February 1998, par. 27.

[13] ECtHR, Matthews v. United Kingdom, No. 24833/94, 18 February 1999, par. 39. The Court started the "living instrument" approach in ECtHR, Tyrer v. United Kingdom, No. 5856/72, 25 April 1978, par. 31.

[14] ECtHR, Christine Goodwin v. United Kingdom, No. 28957/95, 11 July 2002, par. 74; ECtHR, Amann v. Switzerland, No. 27798/95, 16 February 2000, par. 65.

[15] Christine Goodwin v. United Kingdom, No. 28957/95, 11 July 2002, par. 74; ECtHR, Amann v. Switzerland, No. 27798/95, 16 February 2000, par. 65.

[16] Christine Goodwin v. United Kingdom, No. 28957/95, 11 July 2002, par. 74. See also ECtHR, Armonas v.Lithuania, No. 36919/02, 25 November 2008, par. 38.

[17] See Alastair Mowbray, *The creativity of the European Court of Human Rights* (2005) 5(1) HUMAN RIGHTS LAW REVIEW 57.

The European Court of Human Rights says that the mere collection of personal information can interfere with privacy, regardless of how those data are used.[19] If personal information is later disclosed to other authorities, this constitutes a separate interference with privacy.[20]

The European Court of Human Rights has ruled on several cases regarding targeted surveillance,[21] and on two cases on mass surveillance.[22] The Court is very critical of national laws that allow secret surveillance measures. In the Court's words: "in view of the risk that a system of secret surveillance for the protection of national security may undermine or even destroy democracy under the cloak of defending it, the Court must be satisfied that there exist adequate and effective guarantees against abuse."[23]

The mere fact that a country allows mass surveillance interferes with privacy. As the Court puts it:

> [T]he mere existence of legislation which allows a system for the secret monitoring of communications entails a threat of surveillance for all those to whom the legislation may be applied. This threat necessarily strikes at freedom of communication between users of the telecommunications services and thereby amounts in itself to an interference with

---

[18] See on originalism: Bret Boyce, *Originalism and the Fourteenth Amendment*, WAKE FOREST LAW REVIEW, Vol. 33, 1998, 909.

[19] See e.g. ECtHR, Leander v. Sweden, No. 9248/81, 26 March 1987, par. 48; ECtHR, Amann v. Switzerland, No. 27798/95, 16 February 2000, par. 69; ECtHR, Copland v. United Kingdom, No. 62617/00, 3 April 2007, par. 43-44; ECtHR, S. and Marper v. United Kingdom, No. 30562/04 and 30566/04. 4 December 2008, par. 67, par. 121.

[20] ECtHR, Weber and Saravia v. Germany, no. 54934/00, 29 June 2006, par. 79.

[21] See e.g. ECtHR, Klass and others v. Germany, No. 5029/71, 6 September 1978.

[22] ECtHR, Weber and Saravia v. Germany, no. 54934/00, 29 June 2006; ECtHR, Liberty and Others v. the United Kingdom, No. 58243/00, 1 July 2008. The Court speaks of "strategic monitoring" rather than of "mass surveillance" (see Liberty, par. 63; Weber, par. 18).

[23] ECtHR, Weber and Saravia v. Germany, no. 54934/00, 29 June 2006, par. 106.

> the exercise of the (…) rights under Article 8 [privacy], irrespective of any measures actually taken against them.[24]

According to the Court, not only monitoring communications content, but also monitoring metadata interferes with privacy.[25] In sum, the European Court of Human Rights is very critical regarding secret surveillance (see also Part VI). Next, we turn our attention from the Council of Europe and the European Convention on Human Rights to the European Union.

## C. *European Union Charter of Fundamental Rights*

The European Union was created in 1958 to foster economic cooperation between six countries.[26] Since then the EU has grown into an economic and political partnership between 28 European countries.[27] The EU institutions can adopt legal acts, such as directives, regulations, and decisions.[28] A directive is binding, as to the result to be achieved, upon each Member State to which it is addressed, but leaves the choice of form and methods to the national authorities.[29] The Member States must implement directives in their national laws.

The Charter of Fundamental Rights of the European Union lists the fundamental rights and freedoms recognized by the European Union. The Charter was adopted in 2000, and was made a legally binding instrument in 2009.[30] In this paper we use the phrases fundamental rights and human rights interchangeably.[31]

---

[24] ECtHR, Liberty and others v. United Kingdom, No. 58243/00, 1 July 2008, par. 56 (internal citation omitted). See along similar lines (regarding targeted surveillance): ECtHR, Klass and others v. Germany, No. 5029/71, 6 September 1978, par. 37.

[25] ECtHR, Malone v. United Kingdom, No. 8691/79, 2 August 1984, par. 83-84; ECtHR, Copland v. United Kingdom, No. 62617/00, 3 April 2007.

[26] Belgium, France, Italy, Luxemburg, the Netherlands, and West Germany. The Treaty establishing the European Economic Community (the predecessor of the EU) led to the founding of the European Economic Community in 1958 (Treaty Establishing the European Economic Community, 25 March 25, 1957, 298 U.N.T.S. 11)

[27] In 1993, the old name – European Economic Community – was changed to European Union (Treaty on European Union (Maastricht text), 29 July 1992, 1992 O.J. C 191/1).

[28] Article 288 of the Treaty on the Functioning of the European Union.

[29] Article 288 of the Treaty on the Functioning of the European Union.

The Charter copies the right to private life almost verbatim from the European Convention on Human Rights.[32] Article 7 of the Charter says: "[e]veryone has the right to respect for his or her private and family life, home and communications." It follows from the Charter that its article 7 offers at least the same protection as article 8 of the European Convention on Human Rights.[33]

The Charter has a separate provision that lists the limitations that may be imposed on the Charter's rights and freedoms:

> Any limitation on the exercise of the rights and freedoms recognised by this Charter must be provided for by law and respect the essence of those rights and freedoms. Subject to the principle of proportionality, limitations may be made only if they are necessary and genuinely meet objectives of general interest recognised by the Union or the need to protect the rights and freedoms of others.[34]

Regarding the right to private life, the limitations from article 52 of the Charter are similar to those listed in the second paragraph of article 8 of the European Convention on Human Rights.[35]

---

[30] See Article 6(1) of the Treaty on European Union (consolidated version 2012). The institutions of the EU must comply with the Charter. The Member States are also bound to comply with the Charter, when implementing EU law ( 51 of the Charter).

[31] See on the difference: GLORIA GONZÁLEZ FUSTER, *THE EMERGENCE OF PERSONAL DATA PROTECTION AS A FUNDAMENTAL RIGHT OF THE EU (PhD thesis Free University of Brussels)* (Springer 2014), p. 164-166.

[32] The Charter uses the more modern and technology neutral term "communications" instead of "correspondence" (Article 7).

[33] The European Court of Justice says the right to private life in the Charter and the Convention must be interpreted identically. CJEU, C-400/10, J. McB. v L. E., 5 October 2010, par. 53.

[34] Article 52(1) of the Charter of Fundamental Rights of the European Union.

[35] See Article 52(3) of the Charter of Fundamental Rights of the European Union; Note from the Praesidium, Draft Charter of Fundamental Rights of the European Union, doc. no. CHARTE 4473/00, Brussels, 11 October 2000.

In addition to the right to privacy, the European Union Charter of Fundamental Rights contains a separate right to the protection of personal data in article 8, which reads as follows:

> Protection of personal data
>
> 1. Everyone has the right to the protection of personal data concerning him or her.
>
> 2. Such data must be processed fairly for specified purposes and on the basis of the consent of the person concerned or some other legitimate basis laid down by law. Everyone has the right of access to data which has been collected concerning him or her, and the right to have it rectified.
>
> 3. Compliance with these rules shall be subject to control by an independent authority.

The right to data protection in the Charter summarizes some core elements of the Data Protection Directive, which we discuss next.

### D. European Union Data Privacy Directives

In 1995, the EU adopted the Data Protection Directive.[36] One of the aims of the directive is to "protect the fundamental rights and freedoms of natural persons, and in particular their right to privacy with respect to the processing of personal data."[37] The Data Protection Directive became one of the world's most influential data privacy texts.[38]

---

[36] Directive 95/46, of the European Parliament and of the Council of 24 October 1995 on the Protection of Individuals with Regard to the Processing of Personal Data and on the Free Movement of Such Data, 1995 O.J. (L 281) (EC) [hereinafter Data Protection Directive].

[37] Article 1(1) of the Data Protection Directive. The other aim is to safeguard the free flow of personal data between Member States (Article 1(2)).

[38] See Michael Birnhack, *The EU Data Protection Directive: An Engine of a Global Regime*, COMPUTER LAW & SECURITY REVIEW (2008) 24(6) 508; Michael Birnhack, *Reverse Engineering Informational Privacy Law,* 15 YALE JL & TECH. 24 (2012).

Data protection law grants rights to people whose data are being processed (data subjects),[39] and imposes obligations on parties that process personal data (data controllers).[40] Independent Data Protection Authorities oversee compliance with the rules.[41] The Data Protection Directive contains principles for fair data processing, comparable to the Fair Information Practices.[42] Data protection law relies largely on procedural safeguards. The idea is that fair procedures regarding personal data processing should lead to fair outcomes.[43]

De Hert & Gutwirth characterize the right to privacy as an "opacity tool", and data protection law as a "transparency tool."[44] The right to privacy in the European Convention on Human Rights prohibits intrusions into the private sphere.[45] The right to privacy aims to give the individual the chance to remain shielded, or to remain opaque. This prohibition isn't absolute; privacy must often be balanced against other interests, such as the rights of others.

Data protection law takes another approach than the right to privacy, say De Hert & Gutwirth.[46] In principle data protection law allows data processing, if the data controller complies with a number of requirements. Data protection law aims to ensure fairness when personal data are processed.[47] One of the main tools in data protection law to ensure fairness is requiring transparency regarding data processing. Hence: a transparency tool.

---

[39] Article 2(a) of the Data Protection Directive.

[40] Article 2(d) of the Data Protection Directive.

[41] Article 8(3) of the Charter of Fundamental Rights of the European Union; Article 28 of the Data Protection Directive.

[42] The core of the Data Protection Directive isArticle 6. An influential version of the Fair Information Practices is contained in Organisation for Economic Cooperation and Development, *OECD Guidelines on the Protection of Privacy and Transborder Flows of Personal Data*, http://www.oecd.org/sti/ieconomy/oecdguidelinesontheprotectionofprivacyandtransborderflowsofperso naldata.htm, accessed 29 July 2015. See on the Fair Information Practices: Gellman, Robert. "Fair Information Practices: A Basic History" (Version 2.02, November 11, 2013, continuously Updated). http://bobgellman.com/rg-docs/rg-FIPShistory.pdf); Frederik Zuiderveen Borgesius, Jonathan Gray, Mireille Van Eechoud, *Open Data, Privacy, And Fair Information Principles: Towards A Balancing Framework*, BERKELEY TECH. L.J. 2015 (forthcoming)

[43] See COLIN BENNETT, REGULATING PRIVACY: DATA PROTECTION AND PUBLIC POLICY IN EUROPE AND THE UNITED STATES (Cornell University Press 1992), p. 112.

[44] Paul de Hert & Serge Gutwirth, *Privacy, Data Protection and Law Enforcement. Opacity of the Individual and Transparency of Power*, in PRIVACY AND THE CRIMINAL LAW (ERIK CLAES. ANTONY DUFF & SERGE GUTWIRTH EDS. 2006)

[45] See Section II A.

[46] Paul de Hert & Serge Gutwirth, *Privacy, Data Protection and Law Enforcement. Opacity of the Individual and Transparency of Power*, in PRIVACY AND THE CRIMINAL LAW (ERIK CLAES. ANTONY DUFF & SERGE GUTWIRTH EDS. 2006).

[47] SeeArticle 6(1)(a) of the Data Protection Directive.

The e-Privacy Directive[48] complements and specifies the general Data Protection Directive for the telecommunications sector.[49] Most of the e-Privacy Directive's rules only apply to a narrow category of firms: "providers of public communications networks,"[50] and "providers of publicly available electronic communications services."[51] In this paper we use "telecom providers" as shorthand for both provider types. Phone operators and Internet access providers are examples of telecom providers.[52]

The e-Privacy Directive says, in short, that traffic and location data must be erased when they're no longer required for billing or for conveying a communication, unless the user has given consent for another use.[53]

Traffic data are defined in the e-Privacy Directive as "any data processed for the purpose of the conveyance of a communication on an electronic communications network or for the billing thereof."[54] Examples of traffic data are the time of a communication, the email address of communicating partners, and the IP address used to access the internet.[55] Location data are data "indicating the geographic position of the terminal equipment of a user of a publicly available electronic communications service".[56] Location data may show, for instance, the location of a cell phone user.[57] For readability reasons, this paper also speaks of metadata to refer to traffic and location data. In summary, the e-Privacy Directive lays down a strict regime for telecom providers when dealing with metadata.

---

[48] Directive 2002/58, of the European Parliament and of the Council of 22 July 2002 Concerning the Processing of Personal Data and the Protection of Privacy in the Electronic Communications Sector, 2002 O.J. (L 201) 37 (EC) [hereinafter e-Privacy Directive]. The e-Privacy Directive was last amended in 2009, by Directive 2009/136 of the European Parliament and of the Council of 25 November 2009, O.J. (L 337) 11 (EC). In this paper we refer to the consolidated version of the e-Privacy Directive, unless otherwise noted.

[49] Article 1(2) of the e-Privacy Directive.

[50] An electronic communications network is defined in Article 2(a) of the Framework Directive (Council Directive 2002/21 of 7 March 2002 on a common regulatory framework for electronic communications networks and services as amended by Directive 2009/140/EC and Regulation 544/2009 (Framework Directive)).

[51] Article 2(c) of the Framework Directive.

[52] The e-Privacy Directive's background as a directive regulating telecommunications companies can help to explain the narrow scope of many of its provisions. See Axel Arnbak, *Conceptualizing Communications Security: A Value Chain Approach* (Conference paper TPRC 41: THE 41ST RESEARCH CONFERENCE ON COMMUNICATION, INFORMATION AND INTERNET POLICY) (August 15, 2013) <http://ssrn.com/abstract=2242542> accessed 20 July 2015, p. 9. See on the scope of the e-Privacy Directive also Joris van Hoboken & Frederik Zuiderveen Borgesius, Scoping Electronic Communication Privacy Rules: Data, Services and Values, CONFERENCE PAPER EUROCPR 2015 (forthcoming).

[53] Article 6 and 9 of the e-Privacy Directive. See also recital 3 of the Data Retention Directive.

[54] Article 2(b) of the e-Privacy Directive.

But article 15 of the e-Privacy Directive contains an important exception to the strict regime for traffic and location data. Article 15(1) says, in short, that Member States may adopt laws to restrict the e-Privacy Directive's requirements regarding confidentiality of communications and metadata, when such a restriction constitutes a necessary, appropriate, and proportionate measure within a democratic society to safeguard national security, defense, and crime prevention. To this end, Member States may, for instance, adopt data retention obligations. The conditions in Article 15(1) of the e-Privacy Directive resemble Article 8(2) of the European Convention on Human Rights, which states under which conditions states can limit the right to privacy.[58]

The 2006 Data Retention Directive added another exception to the e-Privacy Directive.[59] Since the 2006 amendment, article 15(2) of the e-Privacy Directive says, in short, that article 15(1) does not apply to data that must be retained under the Data Retention Directive.[60] The Data Retention Directive can be seen as one large exception to the metadata privacy regime of the e-Privacy Directive. We discuss the Data Retention Directive in more detail in Section III A.

---

[55] See recital 15 of the e-Privacy Directive, and Article 5 of the Data Retention Directive.
[56] Article 9 of the e-Privacy Directive.
[57] See recital 14 of the e-Privacy Directive.
[58] See WILFRED STEENBRUGGEN, PUBLIEKE DIMENSIES VAN PRIVÉ-COMMUNICATIE: EEN ONDERZOEK NAAR DE VERANTWOORDELIJKHEID VAN DE OVERHEID BIJ DE BESCHERMING VAN VERTROUWELIJKE COMMUNICATIE IN HET DIGITALE TIJDPERK [Public dimensions of private communication: an investigation into the responsibility of the government in the protection of confidential communications in the digital age] (PhD thesis University of Amsterdam) (Academic version 2009), p. 176-190.
[59] Council Directive 2006/24/EC of 15 March 2006 on the retention of data generated or processed in connection with the provision of publicly available electronic communications services or of public communications networks and amending Directive 2002/58/EC (Data Retention Directive).
[60] See on Article 15 of the e-Privacy Directive: Mark Cole and Franziska Boehm, *Data Retention After the Judgement of the Court of Justice of the European Union* (Report for the Greens/EFA Group in the European Parliament) June 2014 http://hdl.handle.net/10993/17500 accessed 27 July 2015, 46-48.

In summary, the e-Privacy Directive lays down a strict and privacy-friendly regime for metadata. Telecom providers must erase or anonymize metadata, unless they are necessary for billing, and only for as long as so necessary. But article 15 makes a hole in that regime. Since the 2006 amendment through the Data Retention Directive, the hole of article 15 has turned into a gaping crater.

### E. Court of Justice of the European Union

The Court of Justice of the European Union is one of the core EU institutions, and is based in Luxemburg.[61] The Court of Justice of the European Union is thus an entirely different entity than the European Court of Human Rights in Strasburg.[62] National judges in the EU can, and in some cases must, ask the Court of Justice of the European Union for a preliminary judgment concerning the validity and interpretation of directives and other acts of the EU.[63]

For a long time, the Court of Justice of the European Union did not feel itself competent to rule on fundamental rights. This can be explained by the fact that the European Union started as an economic community.[64] But in 1969, the Court said "fundamental human rights [are] enshrined in the general principles of Community law and protected by the Court."[65] Now the Treaty on the European Union codifies the importance of human rights.[66] Recently the Court of Justice of the European

---

[61] Article 13(1) of the Treaty of the European Union. http://eur-lex.europa.eu/legal-content/EN/TXT/HTML/?uri=OJ:C:2010:083:FULL&from=EN
[62] See Section II B.
[63] Article 19(3)(b) of the Treaty on European Union. http://eur-lex.europa.eu/legal-content/EN/TXT/?uri=uriserv:OJ.C_.2010.083.01.0001.01.ENG
[64] GLORIA GONZÁLEZ FUSTER, *THE EMERGENCE OF PERSONAL DATA PROTECTION AS A FUNDAMENTAL RIGHT OF THE EU (PhD thesis Free University of Brussels)* (Springer 2014), p. 164.
[65] C-29/69 *Stauder v Stadt Ulm* [1969] ECR 419, Judgment of the Court of 12 November 1969, par. 7. See also C-11/70 Internationale Handelsgesellschaft mbH v Einfuhr- und Vorratsstelle für Getreide und Futtermittel [1970] ECR 1125,
[66] Article 6(3) of the Treaty on the European Union (consolidated version) http://eur-lex.europa.eu/legal-content/EN/TXT/HTML/?uri=OJ:C:2010:083:FULL&from=EN

Union has given influential judgments in the field of data privacy law, such as the Data Retention judgment.[67] It is to this judgment that we turn now.

## III    THE DATA RETENTION DIRECTIVE IS INVALID

### A.  The Data Retention Directive

Since the 1990s, long before the 9/11 terrorist attacks in New York,[68] US and UK authorities have lobbied EU institutions and Member States to adopt data retention legislation. After the terrorist attacks in Madrid (2004) and London (2005), the UK arranged a narrow majority support amongst fellow Member States for a EU-wide data retention requirement. In 2006 it took three months to race the Data Retention Directive through all EU institutions – unusually fast for EU lawmaking.[69]

The Data Retention Directive obliged Member States to adopt metadata retention requirements for telecom providers.[70] Under the directive Member States had to ensure that telecom providers must retain metadata for a period between six months and two years from the date of the communication.[71]

The directive prescribed in detail which types of traffic and location data must be retained. Examples of metadata that had to be retained are data necessary to trace and identify the source of a communication, such as phone numbers.[72] Other examples are the date and time of the log-in and log-off of an Internet access service, together with the IP address and the user ID of the subscriber or registered user.[73] The Data Retention Directive did not require retention of the communication contents; it required the retention of metadata.[74]

---

[67] CJEU, C-293/12 and C-594/12, *Digital Rights Ireland and Seitlinger a.o.*, 8 April 2014. See also CJEU, C-131/12, *Google Spain*, 13 May 2014.

[68] See generally: National Commission on Terrorist Attacks Upon the United States, The 9/11 Commission Report, http://www.9-11commission.gov/report/911Report.pdf, accessed 29 July 2015.

[69] See Marie-Pierre Granger and Kristina Irion, *The Court of Justice and the Data Retention Directive in Digital Rights Ireland: telling off the EU legislator and teaching a lesson in privacy and data protection*, EUROPEAN LAW REVIEW 39, no. 4 (2014), p. 835; Mark Cole and Franziska Boehm, *Data Retention After the Judgement of the Court of Justice of the European Union* (Report for the Greens/EFA Group in the European Parliament) June 2014 http://hdl.handle.net/10993/17500 accessed 27 July 2015, p. 11-12.

[70] Article 3(1) of the Data Retention Directive: "By way of derogation from Articles 5, 6 and 9 of [the e-Privacy] Directive 2002/58/EC, Member States shall adopt measures to ensure that the data specified in Article 5 of this Directive are retained in accordance with the provisions thereof, to the extent that those data are generated or processed by providers of publicly available electronic communications services or of a public communications network within their jurisdiction in the process of supplying the communications services concerned."

[71] Article 6 of the Data Retention Directive: "Member States shall ensure that the categories of data specified in Article 5 are retained for periods of not less than six months and not more than two years from the date of the communication."

The Data Retention Directive did not prescribe under which conditions law enforcement bodies could access the retained metadata. The directive did say that Member States had to adopt measures to ensure that retained metadata were provided only to the competent national authorities in specific cases.[75] Regarding data security, the directive said, in short, that Member States had to ensure that telecom providers complied with certain data security principles.[76]

Many types of communication services are outside the scope of the Data Retention Directive, because the directive is an exception to the E-Privacy Directive and thus only applies to conventional telecom and Internet access providers.[77] For instance, providers of smart phone messaging apps, webmail services, and social network sites fall outside the directive's scope.

The Data Retention Directive was controversial from the start. The European Data Protection Supervisor[78] called the directive "the most privacy invasive instrument ever adopted by the EU in terms of scale and the number of people it affects."[79] Before the proceedings for the Court of Justice of the European Union, several national constitutional courts had already ruled that (parts of) the national implementation laws of the directive violated their national constitutions. This happened amongst others in

---

[72] Article 5(1)(a) of the Data Retention Directive.
[73] Article 5(1)(a) of the Data Retention Directive.
[74] Article 5(2) of the Data Retention Directive. "No data revealing the content of the communication may be retained pursuant to this Directive." See also recital 13 of the Data Retention Directive.
[75] Article 4 of the Data Retention Directive.
[76] Article 7 of the Data Retention Directive. See Part V of this paper.
[77] See section II.D.
[78] The European Data Protection Supervisor (EDPS) is the supervisory authority responsible for monitoring the processing of personal data by the EU institutions and bodies (see Article 41 of Regulation (EC) 45/2001 on personal data processing by the Community institutions and bodies).
[79] European Data Protection Supervisor, *The "moment of truth" for the Data Retention Directive: EDPS demands clear evidence of necessity,* 3 December 2010, http://europa.eu/rapid/press-release_EDPS-10-17_en.htm?locale=en, accessed 25 July 2015.

Bulgaria, Romania, Germany, Cyprus, and the Czech Republic.[80] Against this background the Court of Justice of the European Union had to decide the Data Retention case, which we discuss next.

## B. The Data Retention Judgment

In the 2014 Data Retention judgment, the Court of Justice of the European Union strikes down the Data Retention Directive. The case concerns questions from referring judges in Ireland and in Austria. In essence, the national judges asked the Court of Justice of the European Union in to examine the validity of the Data Retention Directive, in the light of Articles 7, 8 and 11 of the Charter of Fundamental Rights of the European Union.[81] These rights concern privacy (article 7), the protection of personal data (article 8), and freedom of expression (article 11).

The Court of Justice of the European Union invalidates the directive, because the EU lawmaker exceeded the limits imposed by the privacy and data protection rights of the Charter. The Court says it is "not inconceivable" that data retention interferes with people's freedom of expression as well.[82] But, as it invalidates the directive already on other grounds, the Court sees no need to examine the directive in the light of freedom of expression.[83] In this paper we also focus on privacy and data protection rights, rather than on freedom of expression.[84]

---

[80] See for an overview of the national cases: Eleni Kosta, *The Way to Luxemburg: National Court Decisions on the Compatibility of the Data Retention Directive with the Rights to Privacy and Data Protection,* SCRIPTed, (2013) 10:3, 33; Mark Cole and Franziska Boehm, *Data Retention After the Judgement of the Court of Justice of the European Union* (Report for the Greens/EFA Group in the European Parliament) June 2014 http://hdl.handle.net/10993/17500 accessed 27 July 2015, 15-18.
[81] CJEU, C-293/12 and C-594/12, *Digital Rights Ireland and Seitlinger a.o.*, 8 April 2014, par. 23.
[82] *Id*. par. 28.
[83] *Id*. par. 71. The referring courts ask several other questions as well. But since the Court of Justice of the European Union already declares the directive invalid, the Court finds there is no need to answer the other questions.
[84] See on the Data Retention Directive and freedom of expression: Patrick Breyer, *Telecommunications data retention and human rights: the compatibility of blanket traffic data retention with the ECHR* EUROPEAN LAW JOURNAL (2005) 11(3) 365, 373.

# IV    MASS METADATA COLLECTION INTERFERES WITH PRIVACY

Below we discuss the implications of the judgment, focusing on three themes: data collection as a privacy interference (Part IV), data security requirements (Part V), and the proceduralization of surveillance law (Part VI).

## A.  Interference with Privacy and Data Protection Rights

In the data retention judgment, the Court of Justice of the European Union calls the privacy interferences required by the Data Retention Directive "wide-ranging" and "particularly serious."[85] The Court notes that metadata can disclose intimate details about people's lives.[86]

> Those data, taken as a whole, may allow very precise conclusions to be drawn concerning the private lives of the persons whose data has been retained, such as the habits of everyday life, permanent or temporary places of residence, daily or other movements, the activities carried out, the social relationships of those persons and the social environments frequented by them.[87]

In line with the case law of the European Court of Human Rights,[88] the Court of Justice of the European Union distinguishes two separate privacy interferences: storing the data, and law enforcement access to the stored data. First, the mere obligation for telecom providers to retain data interferes with the Charter's right to privacy.[89] The Court recognizes that surveillance can lead to chilling effects: data retention "is likely to generate in the minds of the persons concerned the feeling that their private lives are the subject of constant surveillance."[90]

---

[85] The Court says this about the interference with both privacy (Article 7) and data protection (Article 8 of the Charter). CJEU, C-293/12 and C-594/12, *Digital Rights Ireland and Seitlinger a.o.*, 8 April 2014, par. 37.

[86] *Id*. par. 26. The retained data make it possible to learn, for instance, "to know the frequency of the communications of the subscriber or registered user with certain persons during a given period."

[87] *Id*. par. 27.

[88] See Section II B.

[89] CJEU, C-293/12 and C-594/12, *Digital Rights Ireland and Seitlinger a.o.*, 8 April 2014, par. 34. The data retention obligation "constitutes in itself an interference with the [privacy] rights guaranteed by Article 7 of the Charter."

Second, the Court says the access by law enforcement bodies to the stored data also interferes with privacy.[91] The Court adds that "[t]o establish the existence of an interference with the fundamental right to privacy, it does not matter whether the information on the private lives concerned is sensitive or whether the persons concerned have been inconvenienced in any way."[92]

In the US, because of the third party doctrine, judges would probably not have concluded that access by authorities to data that are stored by telecom providers (private parties) constitutes an interference with privacy. The US Supreme Court has said that people do not have a "legitimate 'expectation of privacy'" in personal information held by third parties.[93]

The Court of Justice of the European Union says that, apart from the privacy interference, the directive also interferes with the Charter's right to the protection of personal data.[94] The Court notes that the Data Retention Directive derogates from the regime of the e-Privacy Directive and the Data Protection Directive, which requires confidentiality of communications metadata.[95]

In sum, the Court finds that the Data Retention Directive entails a far-reaching interference with privacy and data protection rights. Can that interference be justified? That is the question for the next section.

---

[90] *Id*. par. 37.
[91] *Id*. par. 35.  This is in line with case law of the European Court of Human Rights:  ECtHR, Weber and Saravia v. Germany, no. 54934/00, 29 June 2006, par. 79.
[92] CJEU, C-293/12 and C-594/12, *Digital Rights Ireland and Seitlinger a.o*., 8 April 2014, par. 33. See also par. 56-57 jo. Par. 5
[93] United States v. Miller, 425 U.S. 435, 442-43 (1976): "The depositor takes the risk, in revealing his affairs to another, that the information will be conveyed by that person to the Government." See Joel R. Reidenberg, *Data Surveillance State in the United States and Europe, the,* 49 Wake Forest L.Rev. 583 (2014), 588.
[94] CJEU, C-293/12 and C-594/12, *Digital Rights Ireland and Seitlinger a.o*., 8 April 2014, par. 36. See also par. 30.
[95] *Id*. par. 32.

### B. Interference: the Justification

As noted, the Charter of Fundamental Rights of the European Union lists the limitations that may be imposed on the Charter's rights and freedoms. Such limitations "must be provided for by law and respect the essence of those rights and freedoms."[96] In addition, "subject to the principle of proportionality, limitations may be made to those rights and freedoms only if they are necessary and genuinely meet objectives of general interest recognized by the Union or the need to protect the rights and freedoms of others."[97]

According to the Court, the Data Retention Directive does not interfere with the *essence* of data protection and privacy rights. The Court arrives at that conclusion, even though it finds the interference with data protection and privacy "particularly serious."[98] The Court says the directive does not affect the essence of the right to data protection, because, in brief, the directive contains some data security requirements.[99]

The Court says the Data Retention Directive does not affect the essence of the right to privacy either, because the directive requires the retention of metadata rather than of communications content.[100] Hence, the Court suggests that collecting metadata is less intrusive than collecting communications data.[101] We are not convinced by that argument (see Part VI).

Next the Court establishes that the directive satisfies an objective of general interest,[102] because it aims "to contribute to the fight against serious crime and thus, ultimately, to public security."[103] According to the Court, the fight against terrorism and serious crime is, indeed, an objective of general interest.[104]

---

[96] Article 52(1) of the Charter of Fundamental Rights of the European Union.
[97] CJEU, C-293/12 and C-594/12, *Digital Rights Ireland and Seitlinger a.o.*, 8 April 2014, par. 38, summarizing the second sentence of Article 52(1) of the Charter of Fundamental Rights of the European Union.
[98] *Id*. par. 39.
[99] *Id*. par. 40.
[100] *Id*. par. 38.
[101] The Court does note, however: "Those [retained] data, taken as a whole, may allow very precise conclusions to be drawn concerning the private lives of the persons whose data has been retained, such as the habits of everyday life, permanent or temporary places of residence, daily or other movements, the activities carried out, the social relationships of those persons and the social environments frequented by them" (*Id*. par. 27).
[102] *Id*. par. 44.

In the end, the Court strikes down the Data Retention Directive because the interference with privacy and data protection rights is disproportional. The Court sees the interference as disproportional because the directive does not comply with certain, mostly procedural, requirements. We discuss the risk of too much focus on procedures (rather than on substantive safeguards) in Part VI, on the *proceduralization* of surveillance law. Now we first turn to data security requirements that follow from human rights case law.

## V    DATA SECURITY REQUIREMENTS UNDER HUMAN RIGHTS LAW

In this Part we discuss the requirements for data security that follow from human rights case law in Europe, and from the Data Retention Directive in particular. Although the Court of Justice of the European Union invalidates the Data Retention Directive because it entails a disproportionate interference with privacy and data protection rights,[105] the Court also discusses the lack of data security safeguards in the directive.

### A.  The Data Retention Judgment on Data Security

The Court of Justice of the European Union says that the fundamental right to data protection requires effective data security standards, and that the Data Retention Directive fails in this respect.[106]

> [The Data Retention] Directive 2006/24 does not provide for
> sufficient safeguards, as required by [the right to the

---

[103] *Id*. par. 41.
[104] *Id*. par. 41. Furthermore says the Court, Article 6 of the Charter "lays down the right of any person not only to liberty, but also to security."
[105] *Id*. par. 65. See also Mark Cole and Franziska Boehm, *Data Retention After the Judgement of the Court of Justice of the European Union* (Report for the Greens/EFA Group in the European Parliament) June 2014 http://hdl.handle.net/10993/17500 accessed 27 July 2015, 38.
[106] *Id*. par. 66.

> protection of personal data in] Article 8 of the Charter, to
> ensure effective protection of the data retained against the risk
> of abuse and against any unlawful access and use of that
> data.[107]

Subsequently, the Court presents a new analytical framework to assess when the EU lawmaker must adopt legal data security safeguards to comply with the Charter's right to data protection. In the future, the EU lawmaker must consider the following three factors when legislating data security: (i) the quantity of data, (ii) the sensitivity of data, and (iii) the risk of unlawful access to data.[108] In the Court's words:

> Article 7 of [the Data Retention] Directive 2006/24 does not
> lay down rules which are specific and adapted to (i) the vast
> quantity of data whose retention is required by that directive,
> (ii) the sensitive nature of that data and (iii) the risk of
> unlawful access to that data, rules which would serve, in
> particular, to govern the protection and security of the data in
> question in a clear and strict manner in order to ensure their
> full integrity and confidentiality. Furthermore, a specific
> obligation on Member States to establish such rules has also
> not been laid down.[109]

The Court adds that the directive allows metadata to be stored outside the EU. As noted, the Charter of Fundamental Rights of the European Union requires that independent Data Protection Authorities oversee compliance with data protection law.[110] According to the Court, Data Protection Authorities cannot effectively monitor compliance if personal data are stored outside the EU.[111] Hence, the Court calls for the retained data to be stored within the EU to enable independent supervision. In this paper we focus on the data security requirements, rather than on data localization as a safeguard for effective supervision.

---

[107] *Id*. par. 66. The European Court of Human Rights also emphasizes the importance of "safeguards against abuse" ECtHR, Weber and Saravia v. Germany, no. 54934/00, 29 June 2006, par. 116. See Section V B.
[108] *Id*. par. 66.
[109] *Id*. par. 66.
[110] Article 8(3) of the Charter of Fundamental Rights of the European Union. See Section II C..
[111] CJEU, C-293/12 and C-594/12, *Digital Rights Ireland and Seitlinger a.o.*, 8 April 2014, par. 68.

The Court's requirement of ensuring "full integrity and confidentiality" of data uses the same language as a widely used definition of information security, known as the *c.i.a.-triad*. According to this definition, "security" assures the confidentiality, integrity and availability of data, programs, and other system resources for authorized users.[112] The c.i.a. triad was developed academic literature in the early 1970s,[113] and remains influential in systems design and technical standards.

In sum, based on human rights, the Court lays down strict data security requirements for stored metadata and personal data. Moreover, as we discuss in the next Section, the Court does not accept it if data security is influenced too much by economic considerations.

### *B. Security Economics*

The Court suggests that a cost-benefit analysis by private parties does not provide sufficient data security safeguards. According to the Court, the Data Retention "does not ensure that a particularly high level of protection and security is applied by [telecom] providers by means of technical and organisational measures."[114] The Court notes that the directive allows telecom providers to consider "economic considerations when determining the level of security which they apply, as regards the costs of implementing security measures."[115]

Scholars of security economics, an increasingly influential field of information security research, have argued for more than a decade that "systems often fail because the organizations that defend them do not bear the full costs of failure."[116] Security economics scholars note that "economic analysis [is] an extremely powerful tool for engineers and policymakers concerned with information security."[117] Security economics uses incentive-based analysis and economic theory to study various security failures. For instance, security economics has explained the causes of security breaches in industries such online banking, operating systems development, and electronic health.[118] The Court seems to agree with security economics scholars: companies' economic considerations do not necessarily provide sufficient data security.

---

[112] C.P. Pfleeger, 'Data security'. In *ENCYCLOPEDIA OF COMPUTER SCIENCE (4TH ED.)*, ANTHONY RALSTON, EDWIN D. REILLY, AND DAVID HEMMENDINGER (EDS.). Chichester: Wiley 2003, p.504.
[113] J. Saltzer & M. Schroeder, The Protection of Information in Computer Systems, Proceedings of the IEEE 63(9) 1975, p.1278-1308. 'Security Controls for Computer Systems', RAND Report R-609 ("The Ware Report"), 1970. Computer Security Technology Planning Study ("The Anderson Report"), 1972.
[114] CJEU, C-293/12 and C-594/12, *Digital Rights Ireland and Seitlinger a.o.*, 8 April 2014, par. 67.
[115] *Id*. par. 67.

The data security provision in the Data Retention Directive (Article 7) refers to the data security provisions in the e-Privacy Directive[119] and the Data Protection Directive.[120] The Court rules that the combined data security provisions of the three directives fail to provide a sufficient data security.[121] These three provisions embody the core of the current approach to legislating data and communications security in the EU. The three data security provisions illustrate the usual way in which the EU lawmaker legislates data security: the law prescribes security measures that take into account risks and costs.

For instance, the Data Protection Directive's data security provision requires "appropriate technical and organizational measures to protect personal data."[122] The provision adds that parties that process personal data can consider the costs when adopting security measures: "[h]aving regard to the state of the art and the cost of their implementation, such [security] measures shall ensure a level of security appropriate to the risks represented by the processing and the nature of the data to be protected."[123] The data security provision of the e-Privacy Directive has a similar structure. Telecom providers must "take appropriate technical and organisational measures to safeguard security of its services."[124] Similar to the Data Protection Directive, the e-Privacy Directive allows telecom providers to consider the costs of security measures.[125] The Court does not accept such economic considerations to play a major role in securing personal data. Hence, the Court implicitly condemns the entire EU approach to data security.

---

[116] A good overview of security economics scholarship is given in T. Moore, R. Anderson, 2011. Internet Security. In: PEITZ, M., WALDFOGEL, J. (EDS.), *THE OXFORD HANDBOOK OF THE DIGITAL ECONOMY*, Oxford University Press. See: ftp://ftp.deas.harvard.edu/techreports/tr-03-11.pdf
[117] Id., p. 22.
[118] Id.
[119] Article 4 of the e-Privacy Directive.
[120] Article 17 of the Data Protection Directive.
[121] CJEU, C-293/12 and C-594/12, *Digital Rights Ireland and Seitlinger a.o*., 8 April 2014, par. 67.
[122] Article 17(1) of the Data Protection Directive.

Apparently the Court is not impressed by the security work by the Data Retention Experts Group, which consists of stakeholders from government, industry, and law enforcement authorities.[126] The Expert Group published a series of guidance documents to develop a "closer understanding of the term 'data security' in relation to its application in [the Data Retention] Directive 2006/24/EC."[127] The Expert Group published several recommendations to augment security, which are not legally binding. The Expert Group consistently held that the Data Retention Directive does not create new security obligations.[128]

A European Commission Decision established the Expert Group.[129] Such a decision is a delegated act of EU law. The Court now suggests that data security must be safeguarded in the same legislation that regulates data retention. After all, the Court laments the lack of specific rules that safeguard data security in a clear and strict manner in the Data Retention Directive.[130] Hence, it appears that fostering data security through delegated acts or voluntary standards is insufficient to comply with human rights.

---

[123] Article 17(1) of the Data Protection Directive says: "Having regard to the state of the art and the cost of their implementation, such [security] measures shall ensure a level of security appropriate to the risks represented by the processing and the nature of the data to be protected."
[124] Article 4(1) of the e-Privacy Directive.
[125] Article 4(1) of the e-Privacy Directive says: "Having regard to the state of the art and the cost of their implementation, these measures shall ensure a level of security appropriate to the risk presented."
[126] Series A: Guidance Documents, doc. 7, DATRET/EXPGRP (2009) 7, final, 11. Oct. 2010.
[127] Idem.
[128] Idem, p. 2.
[129] Idem, p. 2.
[130] See CJEU, C-293/12 and C-594/12, *Digital Rights Ireland and Seitlinger a.o.*, 8 April 2014, par. 66.

Data security requirements had been mentioned in earlier human rights case law in Europe. For instance, in *I. v Finland*, the European Court of Human Rights holds that a country can fail in its obligations under the European Convention on Human Rights if medical personal data are not adequately secured. "What is required in this connection is practical and effective protection to exclude any possibility of unauthorized access occurring in the first place."[131] Several commentators had already suggested that certain data security requirements follow from human rights.[132]

In conclusion, European courts now require effective data security through legislation. In the Data Retention judgment, the Court of Justice of the European Union elevates data security to a central safeguard under the right to the protection of personal data in the Charter of Fundamental Rights of the European Union.[133] The Court tells the EU lawmaker that vast databases cannot exist without clear and precise legislation that ensures full confidentiality and integrity of personal data, and protection against any unlawful access and use of those data. The EU lawmaker should accept its responsibility to keep EU law in line with human rights requirements.

Take, for instance, recent legislative debates on mandatory retention of the Passenger Name Records (PNR) on air travel of every European citizen.[134] The proposal had been voted down in 2013, but has been tabled again after 2015 terrorist attacks.[135] So far, the European institutions have not specifically considered the new data security requirements of the Data Retention judgment. However, with the mass retention of PNR records, the three factors mentioned by the Court – information quantity, sensitivity and the risks of abuse – are clearly at play. Hence, these proposals contradict the data security requirements from the Data Retention judgment, and seem destined for the EU Court when adopted in their current form.

---

[131] ECtHR, I v. Finland, appl. no. 20511/03, 17 July 2008, par. 38-48.
[132] Paul De Hert, *Accountability and System Responsibility: New Concepts in Data Protection Law and Human Rights Law*, MANAGING PRIVACY THROUGH ACCOUNTABILITY 193 (2012).
[133] The ECtHR had already established a similar result in ECtHR, I v. Finland, appl. no. 20511/03, 17 July 2008, par. par. 37.
[134] Proposal for a directive of the European Parliament and of the Council on the use of Passenger Name Record data for the prevention, detection, investigation and prosecution of terrorist offences and serious crime (COM(2011)0032 – C7-0039/2011 – 2011/0023(COD)). See for the current status of the proposal:
http://www.europarl.europa.eu/oeil/popups/ficheprocedure.do?lang=&reference=2011/0023%28COD%29, accessed 31 July 2015.
[135] See BBC News, *Charlie Hebdo: Gun attack on French magazine kills 12*, http://www.bbc.com/news/world-europe-30710883, accessed 31 July 2015.

In sum, the Data Retention judgment sets much stricter obligations for the EU lawmaker to ensure data security. European courts now require data security through legislation that is practical and effective, and protects the "full integrity and confidentiality" of data, especially when information quantity, sensitivity and abuse risks are at play.[136]

## VI THE PROCEDURALIZATION OF MASS SURVEILLANCE LAW

### A. The Data Retention Judgment: Proportionality Assessment

In this Section, we discuss the risk of *proceduralization* of mass surveillance law. The Court invalidates the directive because, taking all circumstances into account, the Directive entails a disproportionate interference with human rights. First, we summarize why the Court of Justice of the European Union finds the Data Retention directive so disproportionate that it is invalid.

As noted, the Court accepts that the directive serves an objective of general interest: public security.[137] But an objective of general interest is not enough; article 52(1) of the Charter also requires attention to proportionality.[138] In the Court's words, "the principle of proportionality requires that acts of the EU institutions be appropriate for attaining the legitimate objectives pursued by the legislation at issue and do not exceed the limits of what is appropriate and necessary in order to achieve those objectives."[139] The Court says that, because of the "seriousness of the interference" with human rights, the EU lawmaker's discretion is reduced.[140]

---

[136] CJEU, C-293/12 and C-594/12, *Digital Rights Ireland and Seitlinger a.o.*, 8 April 2014, par. 66.
[137] Id., par. 43.
[138] Article 52(1) of the Charter of Fundamental Rights of the European Union: "Subject to the principle of proportionality, limitations may be made only if they are necessary and genuinely meet objectives of general interest (…)."

The Court deals step-by-step with the proportionality test. First: appropriateness. The Court says data retention is a valuable tool for criminal investigations, and therefore appropriate to pursue the directive's objective.[141] The Court admits that several methods of electronic communication fall outside the scope of the directive, limiting its effectiveness.[142] But for the Court such loopholes do not make the directive an inappropriate measure.[143]

The second question of the proportionality test asks whether data retention is "necessary" to achieve the objective pursued.[144] While the fight against terrorism is important, says the Court, that objective "does not, in itself, justify a retention measure such as that established by [the directive] being considered to be necessary for the purpose of that fight."[145] In line with earlier case law, the Court affirms that "derogations and limitations in relation to the protection of personal data must apply only in so far as is *strictly necessary*."[146]

Therefore, says the Court, the directive should have given "clear and precise rules governing the scope and application of the measure in question."[147] Such rules should have limited the risk of abuse and of unlawful access to the data.[148] The Court notes that the directive affects virtually everybody in Europe, and requires the retention of many types of metadata.[149]

---

[139] CJEU, C-293/12 and C-594/12, *Digital Rights Ireland and Seitlinger a.o.*, 8 April 2014, par. 46. Relevant factors for the proportionality test include "the area concerned, the nature of the right at issue guaranteed by the Charter, the nature and seriousness of the interference and the object pursued by the interference" (par 47).
[140] CJEU, C-293/12 and C-594/12, *Digital Rights Ireland and Seitlinger a.o.*, 8 April 2014, par. 47.
[141] *Id*. par. 49.
[142] *Id*. par. 50.
[143] *Id*. par. 50. "Whilst, admittedly, that fact is such as to limit the ability of the data retention measure to attain the objective pursued, it is not, however, such as to make that measure inappropriate."
[144] *Id*. par. 51.
[145] *Id*. par. 51.
[146] *Id*. par. 52 (emphasis added).
[147] *Id*. par. 54.
[148] *Id*. par. 54.
[149] *Id*. par. 56.

> It therefore applies to all means of electronic communication, the use of which is very widespread and of growing importance in people's everyday lives. Furthermore (...), the directive covers all subscribers and registered users. It therefore entails an interference with the fundamental rights of practically the entire European population.[150]

According to the Court, the directive "entails a wide-ranging and particularly serious interference with those fundamental rights in the legal order of the EU, without such an interference being precisely circumscribed by provisions to ensure that it is actually limited to what is strictly necessary."[151]

The Court concludes that the directive goes further than necessary, and gives three arguments.[152] First, the data retention requirement applies to everyone, without exceptions,[153] and even applies to people whose communications are subject to professional secrecy.[154] Moreover, the directive "does not require any relationship between the data whose retention is provided for and a threat to public security."[155] And the directive's retention requirements are not restricted to data regarding a particular period or area, or to crime suspects.[156]

Second, the directive is disproportionate because it does not specify the conditions under which authorities can access the data, and does not specify which crimes are serious enough to justify such access.[157] Moreover, complains the Court, the directive does not require that a court or similar body carry out a prior review if authorities seek access to the retained data.[158] Third, the Court notes that the directive requires

---

[150] *Id*. par. 56. See also the Weber and Saravia v. Germany case, in which the European Court of Human Rights says that one of the "minimum safeguards" for secret surveillance is "a definition of the categories of people liable to have their telephones tapped" in the national law allowing surveillance (ECtHR, Weber and Saravia v. Germany, no. 54934/00, 29 June 2006, par. 95).

[151] CJEU, C-293/12 and C-594/12, *Digital Rights Ireland and Seitlinger a.o.*, 8 April 2014, par. 65.

[152] *Id*. par. 57-68.

[153] *Id*. par. 57.

[154] *Id*. par. 58.

[155] *Id*. par. 59.

[156] *Id*. par. 59.

[157] *Id*. par. 60.

[158] *Id*. par. 62. See on prior judicial review in the context of mass surveillance: TJ McIntyre, Judicial Oversight of Surveillance: The Case of Ireland in Comparative Perspective, in JUDGES AS

that many types of metadata are retained for at least six months, while no distinction is made between types of data, on the basis of how useful they are for the fight against terrorism and serious crime.[159]

The Court concludes that the Data Retention Directive "does not lay down clear and precise rules governing the extent of the interference with the fundamental rights enshrined in Articles 7 [privacy] and 8 [data protection] of the Charter."[160]

In conclusion, the Court invalidates the directive because it entails a disproportionate interference with privacy and data protection rights. The Court does not say that mass metadata retention is always incompatible with human rights.

### B. Proceduralization in Human Rights Case Law

We warn against the risk of the proceduralization of mass surveillance law. While we agree with the end result of the Data Retention judgment – declaring the entire directive void – we do not agree that mass metadata surveillance is acceptable, as long as there are sufficient procedural safeguards.

The arguments of the Court of Justice of the European Union for invalidating the directive are largely *procedural* arguments. Taking a skeptical view, the judgment could be read as a checklist for lawmakers to design a mass data retention obligation that does comply with the Charter's proportionality requirement.

The Court lists several procedural arguments why the Data Retention Directive is a disproportionate measure. In brief: (i) the directive lacks safeguards regarding law enforcement access to data,[161] and (ii) lacks specified retention periods for different metadata types in relation to their usefulness for law enforcement.[162] In addition, the Court complains about (iii) insufficient data security safeguards, and (iv) the lack of a

---

GUARDIANS OF CONSTITUTIONALISM AND HUMAN RIGHTS, EDITED BY MARTIN SCHEININ, HELLE KRUNKE, AND MARINA AKSENOVA. CHELTENHAM: EDWARD ELGAR (2015).
[159] *Id*. par. 63.
[160] *Id*. par. 65.
[161] CJEU, C-293/12 and C-594/12, *Digital Rights Ireland and Seitlinger a.o*., 8 April 2014, par. 61.
[162] *Id*. par. 63.

requirement of data storage within the EU.[163] A fifth reason why the directive is a disproportionate measures is that it affects almost everybody in Europe.

The EU lawmaker could meet most of the Court's complaints by designing a new Data Retention Directive, with better procedural safeguards. The EU lawmaker could design a data retention scheme that includes (i) detailed rules for law enforcement access, (ii) specified retention periods, and requirements regarding (iii) data security and (iv) data localization.

The Court's complaint that the directive affects almost everybody in Europe is a real hurdle against mass surveillance measures. After all, mass surveillance typically concerns millions of people. However, it is unclear whether the fact that almost the whole EU population is affected in itself makes the directive disproportional for the Court. The Court mentions the fact that almost everybody is affected by data retention as *one* of the arguments to invalidate the directive – not as a trump argument.[164]

In sum, we the Court of Justice of the European Union could have taken an even tougher stand on mass metadata surveillance. As noted, the Court did not say that metadata surveillance affects the *essence* of privacy.[165] Furthermore, the Court could have explicitly said that the mere fact that the Data Retention Directive affects almost all Europeans makes the directive disproportionate and therefore invalid.

---

[163] *Id*. par. 66-67 (data security); par. 68 (data localization).

[164] The Court says that the EU lawmaker went too far, "having regard to all the foregoing considerations" (*Id*. par. 69).

[165] The Constitutional Court of Romania says the data retention directive threatens the essence of privacy (Decision of the Romanian Constitutional Court 1258, 08 October 2009. The original decision in Romanian is available at http://www.legi-internet.ro/fileadmin/editor_folder/pdf/Decizie_curtea_constitutionala_pastrarea_datelor_de_trafic.pdf Unofficial translation by Bogdan Manolea and Anca Argesiu, http://www.legi-internet.ro/fileadmin/editor_folder/pdf/decision-constitutional-court-romania-data-retention.pdf). But see Boehm and Cole, who say that it was unlikely the Court would declare that the Data Retention Directive interfered with the "essence" of rights: "the application of this criterion is limited to extreme cases of severe infringements and it does not come as a surprise that the Court rejects an infringement of the essence of both rights." Mark Cole and Franziska Boehm, *Data Retention After the Judgement of the Court of Justice of the European Union* (Report for the Greens/EFA Group in the European Parliament) June 2014 http://hdl.handle.net/10993/17500 accessed 27 July 2015, 33.

In the context of surveillance case law of the European Court of Human Rights, scholars have warned for a long time against "the danger or proceduralisation".[166] For instance, in *Weber and Saravia v. Germany* (2006) the European Court of Human Rights declared a German case on mass surveillance inadmissible (and found no violation with the Convention).[167] In brief, the German laws contained sufficient procedural safeguards: "there existed adequate and effective guarantees against abuses of the State's strategic monitoring powers."[168]

*Liberty* v. *the United Kingdom* (2008) concerned a mass surveillance system in the UK. The UK Ministry of Defence operated an Electronic Test Facility, which intercepted all public telecommunications, including phone and email, carried on microwave radio between two stations, a link which also carried much of Ireland's telecommunications traffic. The European Court of Human Rights said that the UK violated the right to privacy of the European Convention on Human Rights. The Court found that the UK violated the Convention because the surveillance was not "in accordance with the law": the UK surveillance laws were too vague and lacked sufficient safeguards again abuse.[169] But the Court did not discuss the question whether the mass surveillance was "necessary in a democratic society". As De Hert and Gutwirth note, the Court focused on procedures, and sidestepped "the political question whether (processing) power should be limited, stopped or prohibited."[170]

---

[166] Paul de Hert and Serge Gutwirth, *Privacy, Data Protection and Law Enforcement. Opacity of the Individual and Transparency of Power*, in PRIVACY AND THE CRIMINAL LAW (ERIK CLAES. ANTONY DUFF & SERGE GUTWIRTH EDS. 2006).

[167] ECtHR, Weber and Saravia v. Germany, no. 54934/00, 29 June 2006, par. 138: "Accordingly, the applicants' complaints under Article 8 must be dismissed as being manifestly ill-founded, in accordance with Article 35 §§ 3 and 4 of the Convention."

[168] ECtHR, Weber and Saravia v. Germany, no. 54934/00, 29 June 2006, par. 137.

[169] ECtHR, Liberty and others v. United Kingdom, No. 58243/00, 1 July 2008, par. 69-70: "the Court does not consider that the domestic law at the relevant time indicated with sufficient clarity, so as to provide adequate protection against abuse of power, the scope or manner of exercise of the very wide discretion conferred on the State to intercept and examine external communications. In particular, it did not, as required by the Court's case-law, set out in a form accessible to the public any indication of the procedure to be followed for selecting for examination, sharing, storing and destroying intercepted material. The interference with the applicants' rights under Article 8 was not, therefore, 'in accordance with the law'. (…) It follows that there has been a violation of Article 8 in this case."

[170] Paul de Hert and Serge Gutwirth, *Data Protection in the Case Law of Strasbourg and Luxemburg: Constitutionalisation in Action*, in REINVENTING DATA PROTECTION?, EDITED BY S. GUTWIRTH, Y. POULLET, P. DE HERT, C. DE TERWANGNE AND S. NOUWT, 3-44: Springer, 2009, p. 21.

Too much focus on procedural safeguards, note De Hert & Gutwirth, "might well bring the erosion of recognized rights."[171] We agree. And, as we discuss in the next Section, we think mass surveillance is a problem – even with procedural safeguards.

### C. With Procedural Safeguards, Mass Surveillance is Still a Problem

"The hard truth," says a UN report, "is that the use of mass surveillance technology effectively does away with the right to privacy of communications on the Internet altogether. (…) The adoption of mass surveillance technology undoubtedly impinges on the very essence of that right."[172] The report concludes "it is incompatible with existing concepts of privacy for States to collect all communications or metadata all the time indiscriminately. The very essence of the right to the privacy of communication is that infringements must be exceptional, and justified on a case-by-case basis."[173]

Moreover, collecting metadata about virtually everyone is hard to reconcile with the presumption of innocence.[174] As more than a thousand academics contend in their "Academics Against Mass Surveillance" declaration: "mass surveillance turns the presumption of innocence into a presumption of guilt."[175]

---

[171] Paul de Hert & Serge Gutwirth, *Privacy, Data Protection and Law Enforcement. Opacity of the Individual and Transparency of Power*, *in* PRIVACY AND THE CRIMINAL LAW 91 (ERIK CLAES. ANTONY DUFF & SERGE GUTWIRTH EDS. 2006).

[172] UN Special Rapporteur on the promotion and protection of human rights and fundamental freedoms while countering terrorism. 'Report of the Special Rapporteur on the promotion and protection of human rights and fundamental freedoms while countering terrorism (A/69/397, Sixty-ninth session, Agenda item 68 (a))' (23 September 2014) <http://s3.documentcloud.org/documents/1312939/un-report-on-human-rights-and-terrorism.pdf> accessed on 21 October 2014. par. 12 and 18.

[173] UN Special Rapporteur on the promotion and protection of human rights and fundamental freedoms while countering terrorism. 'Report of the Special Rapporteur on the promotion and protection of human rights and fundamental freedoms while countering terrorism (A/69/397, Sixty-ninth session, Agenda item 68 (a))' (23 September 2014) <http://s3.documentcloud.org/documents/1312939/un-report-on-human-rights-and-terrorism.pdf> accessed on 21 October 2014. par. 18.

[174] Paul De Hert, *Balancing Security and Liberty within the European Human Rights Framework. A Critical Reading of the Court's Case Law in the Light of Surveillance and Criminal Law Enforcement Strategies After 9/11*, *UTRECHT L.REV*. 1, (2005): 68: "one could question whether collecting data on the population at large (and not only on suspected persons) is reconcilable with the notion of the presumption of innocence and fairness."

[175] Academics Against Mass Surveillance, 2014, www.academicsagainstsurveillance.org, accessed 29 July 2015. See also Joel R. Reidenberg, *Data Surveillance State in the United States and Europe, the*,

We do not agree with the suggestion of the European Court of Justice that collecting metadata is less intrusive than collecting communications content.[176] First, metadata are easier to analyze than communications content. [177] The contents of communications, such as phone calls or streaming video, are hard to search automatically. But metadata are generally structured, and can easily be searched and analyzed. As computer scientist Felten notes, "[t]he structured nature of metadata makes it easy to analyze massive datasets using sophisticated data-mining and link-analysis programs."[178] Second, collecting metadata (rather than content) enables authorities to capture data about millions of people. Because of data storage costs, it is more difficult (although still possible) to collect communications content on such a large scale. Third, with modern technology, the line between metadata and communications content becomes increasingly fuzzy. Sometimes, notes Felten, "metadata is even more sensitive than the contents of a communication."[179] For instance, metadata can show whether we communicate with a priest, an imam, or Alcoholics Anonymous. Metadata can reveal who our friends, partners or business partners are, and whether people engage in adultery.[180] In sum, as many scholars note, the distinction between communications content and metadata is passé.[181] Metadata are at least as sensitive as communication content.

---

49 WAKE FOREST L.REV. 583 (2014), 605: "the transparency [of citizens resulting from metadata surveillance] reverses the presumption of innocence."

[176] Par. 40. See IV B of this paper.

[177] Edward Felten, *Written Testimony, Committee on the Judiciary Hearing on Continued Oversight of the Foreign Intelligence Surveillance Act*, 2 October 2013, www.cs.princeton.edu/~felten/testimony-2013-10-02.pdf, accessed 29 July 2015, p. 4.

[178] *Id.*, p. 4.

[179] *Id.*, p. 9.

[180] *Id.*, p. 9.

[181] See generally: BERT-JAAP KOOPS AND JAN SMITS, VERKEERSGEGEVENS EN ARTIKEL 13 GRONDWET. EEN TECHNISCHE EN JURIDISCHE ANALYSE VAN HET ONDERSCHEID TUSSEN VERKEERSGEGEVENS EN INHOUD VAN COMMUNICATIE *[Traffic data andArticle 13 of the Constitution. Technical and legal analysis of the distinction between traffic data and communications content]* (Wolf Legal Publishers 2014); JOHAN FISCHER, COMMUNICATIONS NETWORK TRAFFIC DATA - TECHNICAL AND LEGAL ASPECTS (PhD thesis University of Eindhoven), Academic version 2010 <http://alexandria.tue.nl/extra2/689860.pdf> accessed 29 July 2015. (Technische Universiteit Eindhoven); Article 29 Working Party, *Opinion 04/2014 on surveillance of electronic communications for intelligence and national security purposes*, 10 april 2014, p. 5; Patrick Breyer, *Telecommunications data retention and human rights: the compatibility of blanket traffic data retention with the ECHR* EUROPEAN LAW JOURNAL (2005) 11(3) 365, p. 370-371.

Procedural safeguards are important.[182] But we do not think that procedural safeguards are sufficient to regulate mass metadata surveillance. Even with procedural safeguards, mass surveillance is unacceptable.

## VII   THE FUTURE OF MASS DATA RETENTION IN EUROPE

In this Part we discuss the road ahead for mass surveillance in the EU. "Western Europe at the turn of the 21st century" is, as Pinker puts it, "the safest place in human history".[183] Nevertheless, some EU countries find it important to adopt mass surveillance measures.

Blanket metadata retention can still exist in Europe for two reasons. First, the invalidity of the Data Retention Directive does not mean that the national implementing acts automatically became invalid too.[184] With the Data Retention judgment, the legal *obligation* for EU Member States to implement the Data Retention Directive evaporated. But article 15(1) of the e-Privacy Directive still provides a *possibility* for Member States to adopt data retention obligations.[185]

Article 15(1) of the e-Privacy Directive permits national laws that require data retention, but only under strict conditions. Therefore, a national data retention regime must be, for instance, necessary, appropriate, and proportionate. In addition to the requirements of article 15, national data retention laws must observe the requirements of the Data Retention judgment.[186] But under those conditions, national data retention laws are still legally possible.

---

[182] See TJ McIntyre, Judicial Oversight of Surveillance: The Case of Ireland in Comparative Perspective, in JUDGES AS GUARDIANS OF CONSTITUTIONALISM AND HUMAN RIGHTS, EDITED BY MARTIN SCHEININ, HELLE KRUNKE, AND MARINA AKSENOVA. CHELTENHAM: EDWARD ELGAR (2015).

[183] STEVEN PINKER, THE BETTER ANGELS OF OUR NATURE: THE DECLINE OF VIOLENCE IN HISTORY AND ITS CAUSES (Penguin UK (paperback) 2011), p. 62. See also p. 415: The 9/11 attacks killed nearly 3,000 people. Meanwhile, "[e]very year more than 40,000 Americans are killed in traffic accidents, 20,000 in falls, 18,000 in homicides, 3,000 by drowning (including 300 in bathtubs), 3,000 in fires, 24,000 from accidental poisoning, 2,500 from complications of surgery, 300 from suffocation in bed, 300 from inhalation of gastric contents, and 17,000 by "other and unspecified nontransport accidents and their sequelae" (footnote omitted).

[184] In the words of Vandamme: the "invalidation [of a directive] has no *per se* consequences for national implementation law. (…) Thus, any consequences *per se* of the invalidation of a directive hinge upon the national legal systems and the legal authority they attribute to that directive in the course of their domestic legislative processes." THOMAS VANDAMME, THE INVALID DIRECTIVE: THE LEGAL AUTHORITY OF A UNION ACT REQUIRING DOMESTIC LAW MAKING *(PhD thesis University of Amsterdam)* (Europa Law Publishing 2005), p. 326.

[185] See Section III D.

A second reason why blanket metadata retention is still possible in Europe is that the Data Retention judgment has not outlawed such mass surveillance measures per se. Hence, the EU lawmaker could still adopt a new Data Retention Directive, while aiming to comply with the judgment's strict procedural requirements. The Data Retention judgment does not make clear whether the sole fact that mass surveillance affects all Europeans makes such surveillance disproportionate.

At the national level, the response to the judgment has been mixed. The result is a European patchwork for data retention.[187] After the Data Retention judgment, courts in Austria,[188] Slovenia,[189] and Romania[190] quickly annulled the national implementation laws. In Belgium and the Netherlands, judges have invalidated national data retention laws after public interest groups challenged those laws.[191] Conversely, countries such as the UK, still maintain national data retention schemes.[192]

In December 2014, a Member of the European Commission suggested that the Commission had started to work on a new Data Retention Directive. The Commissioner stated that he was examining "how", rather than "if", to re-introduce data retention.[193] After an outburst of critique from public interest groups, the next day a Commission spokesperson hastened to add that the Commission still considered the "if" question as well.[194] In April 2015, the Commission said it plans to launch a consultation on data retention with relevant stakeholders.[195] It seems thus possible that the Commission will propose a new EU data retention obligation.

---

[186] See Article 29 Working Party, *Statement on the ruling of the Court of Justice of the European Union (CJEU) which invalidates the Data Retention Directive* (WP220), Brussels, 1 August 2014, p. 2.
[187] An updated overview is maintained at: http://wiki.vorratsdatenspeicherung.de/Overview_of_national_data_retention_policies, accessed 31 July 2015.
[188] Mark Cole and Franziska Boehm, *Data Retention After the Judgement of the Court of Justice of the European Union* (Report for the Greens/EFA Group in the European Parliament) June 2014 http://hdl.handle.net/10993/17500 accessed 27 July 2015, 55.
[189] *Id.*, 56.
[190] *Id.*, 56.
[191] See EDRi, *Belgian Constitutional Court rules against data retention,* https://edri.org/belgian-constitutional-court-rules-against-dataretention/, accessed 31 July 2015; District Court of The Hague, Verdict in preliminary relief proceedings dated 11 March 2015, ECLI:NL:RBDHA:2015:2498. An unofficial translation by the Interdisciplinary Internet Institute can be found at http://theiii.org/documents/DutchDataRetentionRulinginEnglish.pdf, accessed 31 July 2015.
[192] In the UK, however, the High Court required the UK government to amend the data retention act (DRIPA) to make it comply with European Union Law. David Davis and others v Secretary of State for the Home Department, [2015] EWHC 2092 (Admin), 17 July 2015.

## VIII CONCLUSION

In this paper we discussed the regulation of mass metadata surveillance in Europe. We focused on the Data Retention judgment, in which the Court of Justice of the European Union strikes down the Data Retention Directive that required the retention of metadata of almost all Europeans, to enable access to those data for authorities, in view of fighting serious crime and terrorism. The Court confirms that the mere collection of metadata about people interferes with privacy. In addition, the Court develops three new criteria for assessing the level of data security required from a human rights perspective. Security measures should take into account the risk of unlawful access to data, and the data's quantity and sensitivity. By connecting its ruling to the e-Privacy Directive and the general Data Protection Directive, the Court implicitly condemns the current legal regime for data security in EU law. Hence, the Court instructs the EU lawmaker to take responsibility for robust data security protections in future legislation. As such, the Data Retention judgment is a landmark ruling, which is good news for human rights.

---

[193] Dimitris Avramopoulos, *Exchange of Views between Commissioner Dimitris Avramopoulos and MEPs at the LIBE Committee in the European Parliament*, 3 December 2014, http://europa.eu/rapid/press-release_SPEECH-14-2351_en.htm, accessed 31 July 2015.

[194] "[W]e are now reflecting on the how, rather than the if," said the European Commission to the site Netzpolitik. Netzpolitik, *Telecommunications Data Retention: EU Commission is working on new Data Retention Directive (Update)*, https://netzpolitik.org/2014/suspicionless-mass-surveillance-eu-commission-is-working-on-a-new-data-retention-directive/, accessed 31 July 2015.

[195] Parliamentary questions, 23 April 2015, *Answer given by Mr Avramopoulos on behalf of the Commission*, E 2894/2015: "During the meeting of the Justice and Home Affairs Council which took place in Brussels on 12 March 2015, the Commissioner for Migration, Home Affairs and Citizenship announced that the Commission does not plan to present a new legislative initiative on Data Retention. Instead, the Commission intends to launch a public consultation on the matter with relevant stakeholders."

However, we warn for the risk of proceduralization of mass surveillance law. The Court did not fully condemn mass surveillance that relies on metadata. The judgment leaves open the possibility that, if policymakers lay down sufficient procedural safeguards, mass surveillance can be claimed to be lawful.

So what will happen next in Europe? Even if the European Commission does not present a new Data Retention Directive, Member States can still require data retention. And if a successor of the Data Retention Directive is adopted, it remains to be seen whether the European people still have the energy, stamina, and funding for another 8 years of legal battle. In this light, the Data Retention judgment, while welcomed by human rights groups across the board, may be a Pyrrhic victory in the long run.

Here lies the real danger of the proceduralization of mass surveillance law. While a primary function of human rights is to curtail excessive power of government agencies, in the long run proceduralization might accommodate such power. Because the highest constitutional Court of the European Union did not outlaw mass metadata surveillance, "the most privacy invasive instrument ever adopted by the EU in terms of scale and the number of people it affects",[196] the Data Retention judgment may have set a troubling precedent for the future of privacy in Europe.

* * *

---

[196] European Data Protection Supervisor, *The "moment of truth" for the Data Retention Directive: EDPS demands clear evidence of necessity,* 3 December 2010, http://europa.eu/rapid/press-release_EDPS-10-17_en.htm?locale=en, accessed 25 July 2015.